\documentclass[useAMS,usenatbib]{mn2e}
\usepackage{times}
\usepackage{graphicx}

\title[A Mechanism for the Oxygen and Iron Bimodal Radial Distribution]{A Mechanism for the 
Oxygen and Iron Bimodal Radial Distribution Formation in the Disc of our Galaxy}
\author[Acharova et. al.]
{ I. A. Acharova,$^{1}$\thanks{E-mail:
iaacharova@sfedu.ru (IAA); jacques@astro.iag.usp.br (JRDL); unmishurov@sfedu.ru (YuNM);
bshustov@inasan.ru (BMS); atutukov@inasan.ru (AVT); dwiebe@inasan.ru (DSW)} 
J. R. D. L\'epine, $^{2}$ Yu. N. Mishurov,$^{1,3}$  
B.M.Shustov, $^{4}$ A.V.Tutukov, $^{4}$ 
  \newauthor and D.S. Wiebe $^{4}$\\
$^{1}$ Department of Physics of Cosmos, Southern Federal University, 
5 Zorge, Rostov-on-Don, 344090, Russia \\
$^{2}$ Instituto de Astronomia, Geof\'isica e Ci\^encias 
Atmosf\'ericas, Universidade de S\~ao Paulo, Cidade 
Universit\'aria, S\~ao Paulo, SP, Brazil\\
$^{3}$ Special Astrophysical Observatory of Russian Academy of Sciences, 
N.Arkhyz, Karachaevo-Cherkessia, Russia\\
$^{4}$ Institute of Astronomy of the Russian Academy of Sciences, 
48 Pyatnitskaya St., Moscow, 119017, Russia}

\begin{document}

\date{Accepted 2009 xxxx. Received 2009 xxxx; in original form 2009 xxxx}

\maketitle

\pagerange{\pageref{firstpage}--\pageref{lastpage}} \pubyear{2009}

\label{firstpage}

\begin{abstract}
Recently it has been proposed  that there are two types of SN~Ia progenitors -- short-lived and 
long-lived. On the basis of this idea, we develope a theory of a unified mechanism for the 
formation of the bimodal radial 
distribution of iron and oxygen in the Galactic disc. The underlying cause for the 
formation of the fine structure  of the radial abundance pattern is the influence of spiral 
arms, 
specifically, the combined effect of the corotation resonance and turbulent diffusion. From our
modelling we conclude that to explain  the bimodal radial distributions simultaneously
for oxygen and iron and to obtain approximately equal total iron output from  
different types of supernovae, the mean ejected iron mass per supernova event should be the same as quoted in 
literature if maximum mass of stars, that eject heavy elements, is $50\, M_{\odot}$.
For the upper mass limit of $70\, M_{\odot}$  the production of iron by a supernova II explosion
should be increased by about 1.5 times.
\end{abstract}

\begin {keywords}
Galaxy: abundances -- Galaxy: structure.
\end{keywords}


\section{Introduction} 

Galactic nucleosynthesis studies provide a very important information about the structure and evolution of the Milky Way Galaxy, since chemical abundances accumulate and retain signatures of milestones in its history. This is why this research area is often called the ``cosmic archaeology''. Knowledge of heavy element enrichment in galaxies plays a crucial role in various scientific areas.

One of the most prominent features of the chemical composition both in the Milky Way and other 
galaxies is the metallicity gradient. For a long time, it was widely believed that the radial
 metallicity distribution in galactic discs is described simply by a linear function,   
with the more or less constant 
gradient value being typical for the most part of the disc \citep[see e.g.][]{Wielenetal1996}.

Such an oversimplified belief was first broken by \citet{Shaveretal1983} and 
\citet{Twarogetal1997}. They concluded that the abundance distribution in the disc of the Galaxy 
is not described by a linear function with a constant slope. In a series of papers, 
\citet[][ hereafter A02]{Andrievskyetal2002}, \citet[][ hereafter LKA06]{Lucketal2006} and others, 
have shown that in the Milky Way, oxygen and iron demonstrate a bimodal distribution along the 
galactic radius. In the inner part, from about 4 to 6.6~kpc, the gradient is quite steep  
(for iron it is of the order of --0.13\,dex/kpc), whereas in the outer part, approximately 
from 6.6 to 10.6~kpc, the distribution is plateau-like (the gradient is about --0.03\, dex/kpc, 
which is about 4 times smaller in absolute value than in the inner part), the distributions 
being similar both for oxygen and iron. 

This result has posed a very difficult problem that remained unsolved for several years. Indeed, at the epoch, when most researches believed that the abundance radial distribution is linear, they did not pay much attention to the fact that the gradients show up in observations of different elements. They also did not worry about that oxygen and iron have close radial distribution in the disc of the Galaxy. The reason perhaps was as follows. The radial distributions of both types of SNe, averaged over the azimuthal angle in the galactic plane, are approximately exponential with close radial scales. Hence, we may expect that the abundances, expressed in the logarithmic scale, will be approximately linearly distributed along the galactic radius with close gradient values. But the bimodal radial abundance pattern breaks this oversimplified representation.

In the present paper, we continue the research started by \cite{Mishurovetal02} and \cite{Acharovaetal2005}. In the cited papers and others a theory of formation the bimodal abundance pattern under the influence of spiral arms was developed. The main cause for formation of the bimodal distribution is assumed to be the {\bf corotation} resonance which is located close to the solar galactocentric distance \citep[e.g.][ and papers cited therein]{Lepineetal2001}. In the previous papers we have only considered the enrichment of the interstellar medium (ISM) by oxygen, because the main source of oxygen -- SN II -- are strongly concentrated in spiral arms. Their progenitors are massive stars with extremely short life-time, thus, SN II do not escape from spiral arms where they were born. Hence, oxygen is the most obvious indicator for the arm influence on nucleosynthesis (see details in the papers cited above). Our earlier results have strongly suggested that other models of spiral arms, like recurrent wave patterns with different locations of the corotation resonance during galactic life \citep[e.g.][]{SellwoodBinney2002} or with the corotation in the inner part of the Galaxy \citep{Weinberg1994}, etc., failed to explain the bimodal radial distribution of oxygen in the galactic disc.

Some other explanations have been proposed recently for the decrease in abundances of many 
elements with galactocentric distance as well as for the bimodal structure of the observed 
gradient. These explanations are mostly centered on the radially dependent infall rate 
and/or star formation efficiency 
\citep[see, e.g.,][]{Colavittietal2009,Fuetal2009,Magrinietal2009}. In various models the gradient may have both internal (inside-out disc formation) and external (dilution by metal-poor infall from the halo) origin with respect to the disc. In particular, \citet{Colavittietal2009} concluded that the inside-out disc formation and the density threshold are both necessary ingredients for a model which is supposed to reproduce simultaneously all the disc constraints, including steep gradient in the inner disc and flat gradient in the outer disc. However, in studies cited above the gradient slope changes at 11--12 kpc, while observations indicate that the change occurs closer to the Galactic centre, at around 7~kpc (LKA06).

As it was noticed above, the observed radial distribution of iron is similar to oxygen. Unlike 
oxygen, only $\sim1/3$ of iron is produced by SN II. About $2/3$ of it is synthesized during 
SN Ia explosions \citep{Matteucci2004}. For a long time, precursors of these events were 
considered to be quite old objects with ages of the order of several (4--5) billion years. 
Even if they are born in spiral arms, for a long time prior to the explosion they ``forget'' the initial affinity to spiral arms due to both azimuthal and radial migration. So, at first sight, we should not expect strong manifestations of galactic 
spiral arms influence on radial distribution of iron. Hence, the similarity of bimodal radial 
distributions of oxygen and iron in the disc of the Galaxy represents a difficult enigma.

Recently, \citet{Mannuccietal2006} have shown that there may exist two sub-populations of 
SN\,Ia. Events of the first type (short-lived progenitors, 
they were called ``prompt'') have ages less than $10^8$ years, while 
events of the second type (``tardy'') have long-lived progenitors, with delays between formation 
and explosion being measured in billions years (the theoretical basis for such kind of 
division see in \citet{Iben84}).
The frequencies of explosions and iron outputs for both populations are approximately equal. 
Prompt SNe Ia precursors, being quite massive stars, retain their relation to the spiral pattern. Taking into account that SN II and short-lived SN Ia jointly produce $\sim2/3$ of iron, this new 
result opens an opportunity for solving the above problem in the framework of the theory of 
spiral arms influence on radial abundance distribution.

In this paper, we propose a unified mechanism for formation of bimodal radial 
distribution both for oxygen and iron.


\section{Observational data}

\begin{figure}
\includegraphics {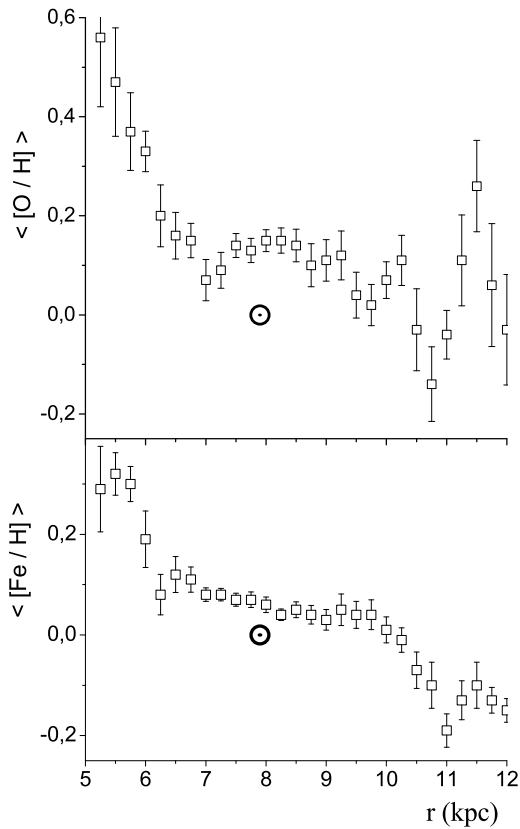}
\caption{ Radial distributions of oxygen (upper panel) and iron (low panel) averaged over the 
azimuthal angle in the galactic plane. 
{\it Open squares} are 
the observed values from \citet{Andrievskyetal2002} and \citet{Lucketal2006} averaged within bins of 
0.5 kpc width. The error-like bars are the standard deviations of the corresponding mean 
values. The abundances of A02 and LKA06 were corrected for the new solar composition according to 
\citet{Asplundetal2006}. Position of the Sun on the diagrams is shown by the usual solar mark.}
\label{fig1}
\end{figure}

In Fig.~\ref{fig1} the radial distribution of $\langle [O/H] \rangle$ and $\langle [Fe/H]\rangle$ 
vs galactocentric distance $r$ is shown (as usual $[X/H]=\log(N_X/N_H)_{star}-
\log(N_X/N_H)_{\odot}$, where $N_{X,H}$ is the number of atoms for $X$ element 
or for hydrogen, correspondingly). We divided $r$ into bins of 0.5~kpc and 
averaged the corresponding data from A02 and LKA06 within each bin. The angle brackets 
$(\langle \dots \rangle)$  denote these averaged values. The error bars show the standard 
deviation of the averaged values within the bin. We used the solar galactocentric distance $r_{\odot}=7.9$ kpc, as in 
LKA06, but abundances were corrected for the new solar composition
\citep{Asplundetal2006}. In this figure, the Sun is shown by the usual mark. The similar bimodal patterns in oxygen and iron radial 
distribution are clearly seen in Fig.~\ref{fig1}. 

Our aim is to explain the bimodal radial distribution of oxygen and iron in the most part of the 
Galaxy, i.e. the steep gradient in the inner part of the Galactic disc ($r\le 7$ kpc) and rather 
shallow distribution up to about 10 kpc with a bending at approximately 7 kpc. The 
peculiarities in the distributions at $r > 10$~kpc are beyond the scope of our theory.


\section{Basic equations and model}

The equations of chemical evolution of ISM are close to the ones of \citet{Tinsley80}, \citet{Chiosi80}, \citet{LaceyFall85}: 
\begin{equation}
\dot{\mu}_g=f-\psi + \int\limits_{m_L}^{m_U}{(m-m_w)\psi(t-\tau_m)\phi (m)\,dm},
\end{equation}

\begin{eqnarray}
{\dot{\mu}_i}&=&\int\limits_{m_L}^{m_U}{(m-m_w)\,Z_i(t-\tau_m)\psi(t-\tau_m)\phi (m)\,dm} + \nonumber\\
&&E_i^{\rm Ia}+E_i^{\rm II} + fZ_{i,f}+\frac {1}{r}\frac {\partial {}}{\partial r}\left 
(r\mu_g D\frac {\partial Z_i}{\partial r}\right),
\end{eqnarray}  
where $\mu_{g,i}$  are surface mass densities for interstellar gas (subscript ``g'') and heavy 
elements (where subscript ``$i$'' is ``O'' for 
oxygen and ``Fe'' for iron), $Z_i=\mu_i/\mu_g$ is the mass fraction for the corresponding 
elements, $\psi$ is the star formation rate (SFR), for which we adopt \citep{Kennicutt98}:
\begin{equation}
\psi=\nu\mu_g^{1.5},
\end{equation}
where $\nu$ is a normalizing coefficient. All the above quantities are functions of time $t$ and radius $r$, but we do not explicitly indicate them except for the cases when there is the time shift, $t-\tau_m$, in the integrals. $m$ is the stellar mass (all masses are in solar units), $m_L$ and $m_U$ are the minimum and maximum stellar masses, $m_L=0.1$, while for $m_U$ we use two values $50$ and $70$ 
\citep{Tsujimotoetal1995}. Our calculations show that the results are very close for the both above values of $m_U$. That is why in our figures we only show the radial distributions for $m_U = 70$.
The life-time of a star of mass $m$, $\tau_m$ (in Gyrs), 
on the main sequence is given by $\log(\tau_m)=0.9-3.8\log(m)+\log^{2}(m)$ \citep{TutukovKruegel1980}. 
The initial mass function $\phi(m)$ is Salpeter's one with the exponent of -~2.35.

In full formulation of the problem, we should take into account the divergent terms, which describe the influence of radial gas flow (``radial inflow'') within the galactic disc. Its effect was considered in various models by \citet{LaceyFall85}, \citet{PortinariChiosi00}, \citet{Mishurovetal02}. In the present paper, we will not include the radial redistribution of the interstellar gas. This problem will be separately considered in details elsewhere.

The infall rate, $f$, of intergalactic gas onto the galactic disc we describe as:
$$f=B(r)\exp(-\frac{t}{t_f}).$$

For the time scale $t_f$ the following representations were used: 
1)~$t_f = 3$~Gyrs (let us call 
this law as ``rapid'' disk formation model, see \cite{PortinariChiosi1999}); 
2)~$t_f = 7$~Gyrs 
(``slow'' disc formation) and 
3)~$t_f = c + d r$, where $c$ and $d$ are some constants. The last dependence produces the 
so-called ``inside-out'' scenario for the galactic disk formation.
We consider various values for $c$ and $d$. {\it i})~$c=1.033$; $d = - 1.267$ 
\citep{Chiappinietal01}. It is obvious that this representation may not be used at 
$r = 1.226$ kpc since here $f \rightarrow 0$. To exclude this peculiarity we use the above 
representation for $r > r_f = 2 –- 3$ kpc. Inside $r_f$ let us suppose $t_f = a + b r_f = const$. 
Our experiments show that in the range of galactocentric distances, interesting for us, 
the result does not depend on the exact value of $r_f$. 
{\it ii})~ $t_f = 1+0.76 r$ \citep{Fuetal2009}. At the Sun location the both expressions 
give $t_f = 7$~Gyrs.

To fix $B(r)$ as usually we adopt that the final radial total density distribution 
(stars + gas) at current epoch, $t=T_D$ ($T_D = 10$ Gyrs is the disc age) is exponential with 
the radial scale $r_d = 2.5 –- 4.5$ kpc (see, e.g., \citet{Naab06}, \citet{Fuetal2009}, 
\citet{SchonBinney08} and others). 

In all our experiments we normalize the infall rate so that the total present—-day density at 
the solar location to be equal to $50 \, M_{\odot}$ \citep{Naab06}. To obtain the normalizing 
coefficient $\nu$ for SFR function (see Equation (3)) we suppose that at the solar galactocentric distance $\mu_g \approx 10 \, M_{\odot}$.

The infalling gas is assumed to have abundances $Z_{i,f}=0.02\,Z_{i,\odot}$.

The value for the mass of stellar remnants (white dwarfs,  
neutron stars and black holes) is adopted as follows: 
for $m \le 10$ $m_w = 0.65m^{0.333}$; in the range  $10 < m < 30$ $m_w = 1.4$; 
if $30 \le m < m_U$ the remnant is a black hole with $m_w = 10$; finally for 
$m \ge m_U$ the stars are black holes right away from their birth and they are removed from the 
nucleosynthesis and returning the mass to ISM (\citet{MassevichTutukov88}, \citet{Breton09},  
\citet{Woosley02}).

The last term in Equation (2) describes the radial diffusion of heavy elements. We need it to 
smooth out strong depression in abundance distribution forming when the role of spiral arms on 
heavy elements 
synthesis is taken into account (in the case of low gradient, the role of the diffusion happens 
to be negligible). Note as well, that the diffusion smoothes out the element 
distribution as admixtures but not the gaseous density as a whole (\citet{LandauLifshitz86}).
The nature of the diffusion we attribute to turbulence in ISM arising due to stellar winds, 
supernovae explosions, galactic fountains, etc. For a phenomenological description of this 
process, we introduce the diffusion term into Equation (2). 
To estimate the diffusion coefficient $D$ we model the turbulent ISM by a system of clouds, 
so a simple gas kinetic approach enables us to derive the expression for $D$
\citep[see ][]{Mishurovetal02, Acharovaetal2005}:
$$D=\frac{Hv_T}{3\pi\sqrt{2}\mu_g}\frac{m_c}{a^2},$$
where $H = 130$~pc is the thickness of the gaseous disc, $v_T = 6.6$~km~s$^{-1}$ is the one dimensional dispersion of cloud chaotic velocity, $m_c$ and $a$ are the mass and radius of a typical cloud, $m_c/a^2 = 100$~M$_\odot$~pc$^{-2}$ \citep{Elmegreen87}.

Enrichment rates of ISM by SN\,Ia and SN\,II, correspondingly, are
set by values $E_i^{\rm Ia}$ and $E_i^{\rm II}$. 
The rates of oxygen and iron synthesis in SN~II events are described by similar expressions 
$E_i^{\rm II}=\eta P_i^{\rm II}R^{\rm II}$, where $P_i^{\rm II}$ is the mass of
ejected oxygen or iron per one SN\,II explosion, $R^{\rm II}$ is the rate of SNe\,II events 
$$R^{\rm II}(r,t)=(1-A)\int\limits_{m_8}^{m_U}{\psi(r,\,t-\tau_m)\phi (m)\,dm},$$ 
$m_8=8$, $A=0.0025$. The last parameter gives the fraction of binary systems producing SNe\,Ia relative to the 
whole mass range $m_L \le m \le m_U$ \citep{Mannuccietal2006}. The factor $\eta$  describes the influence of spiral arms (see below).

Contribution from SN~Ia to the enrichment is represented as $E_i^{\rm Ia}=E_{i,\rm P}+E_{i,\rm T}$  since there are two populations 
of SNe~Ia progenitors, namely, short-lived (``P'') and long-lived \citep[``T'',][]{Mannuccietal2006,Matteuccietal2006}.
As we wrote in Introduction, being massive stars, the 
short-lived progenitors are concentrated near spiral arms. So, by analogy with SN II rates we write down 
$E_{i,\rm P}=\eta\, \gamma\,P_i^{\rm Ia}R_P^{\rm Ia}$ , where $\gamma$  is a correction factor. For 
long-lived SN~Ia  $E_{i,\rm T}=\zeta\,P_i^{\rm Ia}R_T^{\rm Ia}$ . Here $R_{P,T}^{\rm Ia}$  are rates of
SN~Ia explosions for P-
and T-progenitors, $P_i^{\rm Ia}$  is the mass of ejected {\it i-th} element per 
one SN~Ia event, and
$\zeta$  is another normalizing coefficient. 

To compute $R_{P,T}^{\rm Ia}$  we follow \citet{Matteuccietal2006} but explicitly 
separate the ``Delay Time Distribution'' function, $DTD(\tau)$, into two parts: 
$DTD=D_P$  for $\tau \le \tau_S = 10^{-1.07}$ Gyrs (P-progenitors): 
$$\log(D_P)=1.4-50{[\log(\tau)+1.3]}^2,$$ 
and  $DTD=D_T$   for  $\tau>\tau_S$ (T-progenitors): 
$$\log(D_T)=-0.8-0.9{[\log(\tau)+0.3]}^2.$$
So, 
$$R_P^{\rm Ia}(r,t)=A k_{\alpha}\int\limits_{\tau_8}^{\tau_S}{\psi(r,\,t-\tau)D_P(\tau)d\tau},$$ 
$\tau_8$ is the life-time for a star of mass $m=8$ and  
$$R_T^{\rm Ia}(r,t)=Ak_{\alpha}\int\limits_{\tau_S}^{t}{\psi(r,\,t-\tau)D_T(\tau)d\tau},$$ 
$\it{k_{\alpha}}=\int\limits_{m_L}^{m_U}{\phi (m)\,dm}$ \citep[see details in][]{Matteuccietal2006}.

Notice that $D_P$ has a very sharp maximum at $\tau = \tau_P = 10^{-1.3}\, {\rm Gyrs} \approx 50\,$Myrs. This means that for $\tau_8 \le t < \tau_P$ the main contribution to $R_P^{\rm Ia}$ comes from the vicinity of $\tau \sim t$ whereas for $t \ge \tau_P$ the vicinity of $\tau_P$ is most important. The corresponding integral was computed asymptotically using the Laplace method.

To incorporate the influence of spiral arms on formation of radial abundance pattern in our theory, we follow the idea of \citet{Oort1974}. For this, we take into account that SN\,II and short-lived SN\,Ia progenitors are concentrated near spiral arms. Hence, the enrichment of any volume of ISM by heavy elements only takes place when the volume is close to the nucleosynthesis sites, i.~e., inside or near spiral arms. The more frequently the volume passes an arm, the higher the rate of its enrichment gets. The frequency of the volume entering a spiral arm is proportional to the difference $|\Omega(r)-\Omega_P|$  where $\Omega(r)$  is the angular rotation velocity of the galactic disc, $\Omega_P$  is the rotation velocity of the wave pattern responsible for the spiral arms. So, we write $\eta=\beta |\Omega(r)-\Omega_P|\Theta$, where $\beta$  is a normalizing coefficient, $\Theta$ is a cutoff factor. The correction factor $\gamma$  in expression for $E_{i,\rm P}$ was introduced to take into account that we aim to obtain close amounts of iron produced by SN\,II and prompt SN\,Ia.

We assume that long-lived SN\,Ia progenitors do not concentrate in spiral arms and are uniformly distributed over the azimuth angle in the galactic disc. Therefore, unlike $\eta$, $\zeta=const$. 

Spiral arms exist between inner, $r_i$, and outer, $r_o$, Lindblad resonances, their locations being given by the 
condition $\Omega(r_i)-\kappa(r_i)/n \le \Omega_P \le \Omega(r_o)+\kappa(r_o)/n$. Here, $n$ is the number of spiral 
arms, $\kappa$ is the epicyclic frequency. So  in the wave zone, i.~e., between Lindblad resonances, $\Theta=1$, and  beyond them
$\Theta=0$. From the galactic density wave theory \citep{Linetal1969} it is 
known that the density wave pattern 
rotates as a solid body, i.~e., $\Omega_P=const$  whereas the galactic matter rotates differentially: $\Omega$  is a function of the galactocentric distance $r$. The radius $r_c$  where both velocities coincide ($\Omega(r_c)=\Omega_P$) is called the corotation radius. In the absence of diffusion from the above representation for $\eta$ one can expect the formation of a valley (or a gap) in the abundances distribution near the corotation (see below). The diffusion smoothes out the valley so that the combine effect of it and the corotation resonance results in formation of the bimodal radial distribution of the abundances.

\begin{figure}
\includegraphics {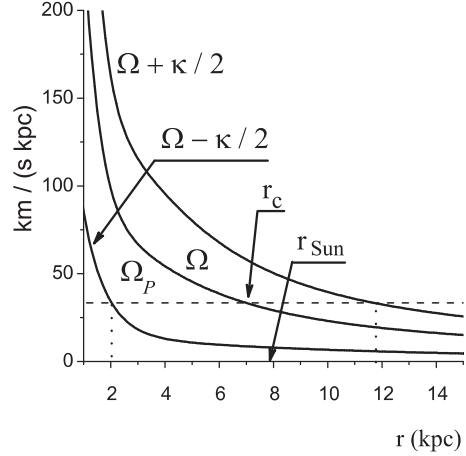}
\caption{The rotation curve $\Omega(r)$ and its combinations with epicyclic 
frequency $\kappa$ approximated over the data of \citet{Clemens1985} adjusted for 
$r_{\odot}=7.9$ kpc. 
The horizontal dashed line shows the adopted value of the angular rotation velocity, 
$\Omega_P$, for the density wave pattern 
which 
corresponds to the location of the corotation resonance at  $r_c \approx 7 \, kpc$. 
The vertical dotted lines indicate the positions of 
Lindblad resonances.}
\label{fig2}
\end{figure}

In our modelling we used a rotation curve based on CO data \citep{Clemens1985} but adjusted for $r_{\odot}=7.9 \,{\rm kpc}$:
\begin{equation}
r\Omega = 260\cdot \exp[-\frac{r}{150}-(\frac{3.6}{r})^2] +  
              360\cdot \exp(-\frac{r}{3.3}-\frac{0.1}{r})\nonumber.  
\end{equation}
The rotation curve is shown in Fig.~\ref{fig2}. Also indicated in Fig.~\ref{fig2} are locations of Lindblad and corotation resonances derived for 
$\Omega_P=33.3$~km\,s$^{-1}$ kpc$^{-1}$  (correspondingly $r_c\approx 7$ kpc). 

Mathematically our system of equations splits into two groups. Equation (1) is an ordinary 
differential equation. 
Solving it, we find both the gaseous and stellar density radial profiles at any time $t$, fitting the observed stellar and gaseous densities to the ones at present time. 
Equations (2) are differential equations in partial derivatives. For the 
boundary conditions we adopt the natural ones at the galactic centre and at the galactic 
periphery which guarantees the finiteness of our solutions 
\citep[see details in][]{Acharovaetal2005}. 

Corresponding constants for ``chemical reactions'' ($\beta, \gamma, \zeta$) were derived in the 
following way. 
As can be seen from \citet{Matteucci2004}, \citet{Tsujimotoetal1995} SNe Ia produce a small part of oxygen, no more than several percents. Hence, at first step we can neglect by oxygen 
enrichment from SNe Ia. This enables us to find independently the parameter $\beta$ by fitting 
the observed radial distribution of oxygen. After that we derive other two quantities, 
$\gamma$ and $\zeta$. To do that we impose the additional condition that short-lived (P) 
SN Ia, long-lived (T) SN Ia, and SN II produce equal amounts of iron ($\sim 1/3$ each). 
This constraint enables to solve the problem unambiguously. The above derived values for 
chemical reactions constants we consider as zero order approximation for them. After that, 
we include the contribution to oxygen production from SN Ia and find the revised chemical 
constants.


\section {Results and discussion}

Below we show the results of our modeling for the adopted value of $\Omega_P$ which corresponds 
to the location of the corotation resonance at $r_c \approx 7$~kpc. 
\footnote{The above value for the corotation radius is not to be considered as a determination 
of $r_c$, it only produces a good fit to the bimodal radial distribution of oxygen and iron in 
a simple model that does not include all the fine effects.}

\begin{figure}
\includegraphics {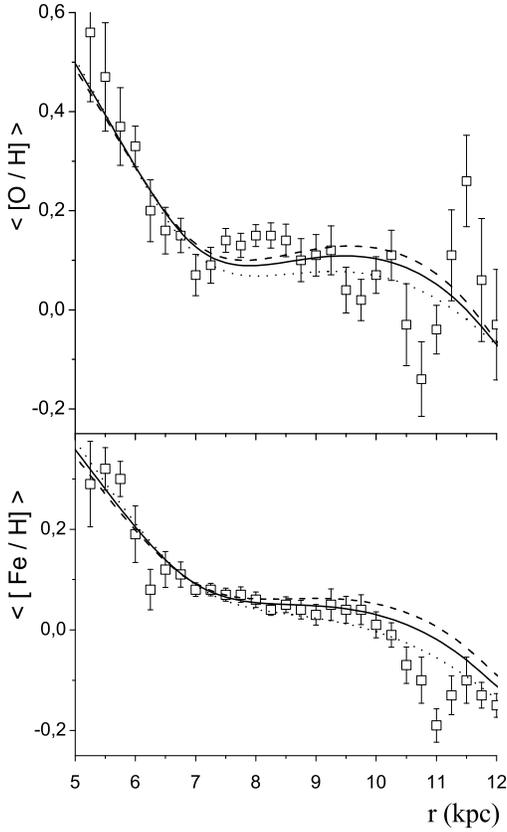}
\caption{Theoretical abundance patterns derived for constant $t_f = 3$ Gyrs 
(rapid disc formation model) superimposed on the 
observations. {\it Dotted lines} are for $r_d =2.5$ kpc. {\it Solid lines} -- for $r_d = 3.5$ kpc.  
{\it Dashed lines} -- for $r_d = 4.5$ kpc.}
\label{fig3}
\end{figure}

\begin{figure}
\includegraphics {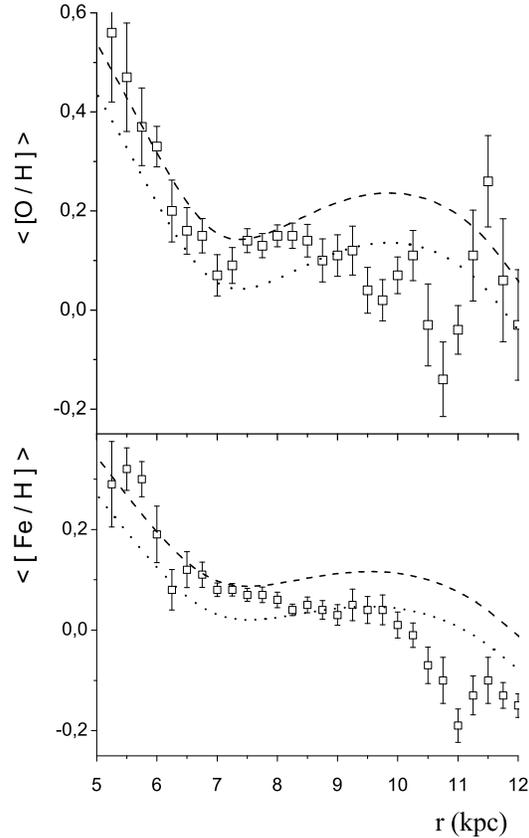}
\caption{ The same as in Fig. 3 but for $t_f=7$ Gyrs (slow disc formation model) and $r_d = 3.5$ 
kpc. {\it Dotted lines} are for the set of the fitting parameters 
$\beta = 0.030$, $\gamma=0.32$, $\zeta=0.22$.
{\it Dashed lines} are for the set 
$\beta = 0.037$, $\gamma=0.30$, $\zeta=0.25$. From this figure is seen that there is no way to 
reconcile the model with the observational data in all range of galactic radius 
by means of the same set of chemical constants.}
\label{fig4}
\end{figure}

In Fig.~\ref{fig3} and Fig.~\ref{fig4} 
are shown the computed abundance patterns for constant 
$t_f$. In the case of rapid disc formation ($t_f = 3$ Gyrs) we demonstrate the dependence of the 
distributions on the radial scale $r_d$. It is seen that in the inner part of the disc the 
corresponding curves are indistinguishable. They slightly differ beyond the solar location. 
Overall, the coincidence of the model with the observations is good but perhaps it is the best 
for the radial scale $r_d = 3.5$ kpc.

In the case of slow disc formation ($t_f = 7$ Gyrs, Fig.~\ref{fig4}) we failed to fit 
the model to the observations. 
The matter is, by means of one set of chemical constants we can fit the observations, say, in 
the inner part of the disc, but for matching with the data in the outer part we have to use 
another set of the constants. 

The general conclusion is: in the framework of scenario of disc formation with constant $t_f$ we 
can succeed in agreement with the observed radial distributions both for oxygen and iron in the 
galactic disc with short time scale $t_f \sim 3$ Gyrs. The fitted parameters happen to be as follows:
for the upper stellar mass $m_U=70$ $\beta=0.019$ , $\gamma=0.44$, $\zeta=0.18$; 
for $m_U=50$ $\beta=0.026$, $\gamma=0.32$, parameter $\zeta$ remains the same. 

However, to derive the observed distributions and satisfy the condition that the above three types of sources 
produce approximately equal amount 
of iron, we had to adopt the following masses ejected per one SN. In the case $m_U=50$ the corresponding 
values are equal to the ones in \citet{Tsujimotoetal1995}, i.~e., 
$P_{\rm O}^{\rm II}=1.80\,M_{\odot}$,
$P_{\rm Fe}^{\rm Ia}=0.613\,M_{\odot}$, $P_{\rm Fe}^{\rm II}=0.084\,M_{\odot}$. 
For the upper value of $m_U\, =\, 70$ 
$P_{\rm O}^{\rm II}=2.47\,M_{\odot}$,
$P_{\rm Fe}^{\rm Ia}=0.613\,M_{\odot}$, like in the last cited paper but 
$P_{\rm Fe}^{\rm II}=0.13\,M_{\odot}$,   
\citep[i.e. about 1.5 times greater than in][]{Tsujimotoetal1995}. 

Our experiments show that SNIa produce about 2\% of oxygen. 

\begin{figure}
\includegraphics {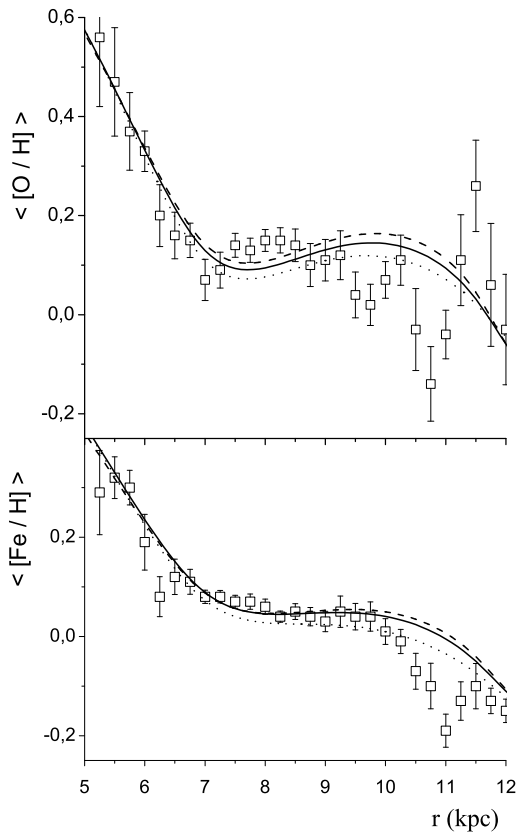}
\caption{ The same as in Fig. 3 but for inside-out disc formation scenario for Chiappini -- like 
model (see Section 3) and $r_f=3$ kpc (in the radius range interesting for us the results 
slightly depend on the exact value of $r_f$). Other designations are the same as in Fig. 3. }
\label{fig5}
\end{figure}

In Fig.~\ref{fig5} are shown the computed distributions for the inside-out scenario 
(since they are close both for the models of \citet{Chiappinietal01} and  \citet{Fuetal2009} we only show the results for Chiappini-like representation of $t_f$). In our opinion the coincidence of the theory with the observations both for oxygen and iron is excellent. The chemical reactions are close to the ones of constant $t_f$, namely: 
for $m_U=70$ $\beta=0.021$ , $\gamma=0.46$, $\zeta=0.18$; 
for $m_U=50$ $\beta=0.032$, $\gamma=0.30$, parameter $\zeta$ again remains the same.

\begin{figure}
\includegraphics {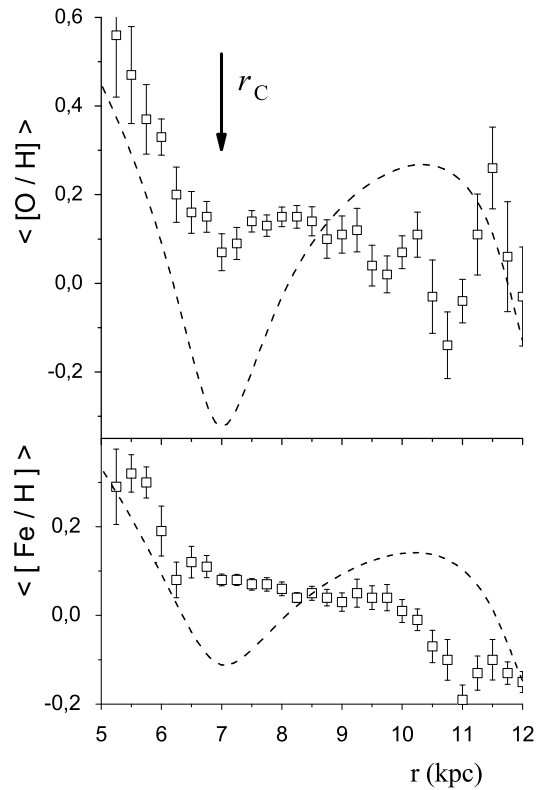}
\caption{ The same as in Fig. 3 but for the diffusion coefficient 10 times less and $r_d=3.5$ kpc. 
By the arrow is shown the location of the corotation resonance. The formation of the deep gap 
in abundances distributions near the corotation in the absence of the turbulent diffusion is 
obvious.
}
\label{fig6}
\end{figure}

To feel the role of the turbulent diffusion, in Fig.~\ref{fig6} we show the theoretical radial 
distributions in the case when the diffusion coefficient is 10 times less (rather arbitrary) 
than the value estimated for the parameters from Section 3 (we can not put it zero since the 
diffusion term in Equation (2) determines the highest order of the derivative). The formation of a gap 
in the abundance radial distribution near the corotation in this case is obvious.

So, as it might have been expected, the main driver for the origin of the bimodal radial 
abundance distribution is the corotation resonance. However, in the absence of the turbulent 
diffusion we will have strong depression in radial distribution of heavy elements near the 
corotation. Hence, only combined effect of corotation and 
diffusion leads to a formation of the bimodal distributions both for oxygen and 
iron. The gradient in the inner part derived from iron is slightly shallower then the one for 
oxygen since $\sim 1/3$ of iron is synthesized by long-lived SNe\,Ia which are not concentrated 
in spiral arms and are not affected by the corotation resonance.

In our theory, we aimed to construct a simplest model for explanation of bimodal radial abundance pattern formation for elements, which are synthesized by different sources with different galactic distributions. That is why we did not take into account various secondary effects like radial inflow of interstellar gas, dependence of stellar life-time on the metallicity, drift 
of the corotation resonance,~etc. Inclusion of such effects in our model will enable to develop a more precise model of formation of the bimodal distribution of heavy elements. Also, it should be noted that in the current study the proposed mechanism acts together with the radially dependent gas infall. However, it will, probably, shape the heavy element distribution in a similar manner acting in concord with other gradient formation mechanisms, like the radially dependent enriched galactic wind, as proposed by \citet{Wiebeetal2001}. Also, as \citet{Acharovaetal2005} have shown, the influence of spiral arms alone is able to explain the very existence of the gradient, not only its bimodal shape. In any case, the 
discovery of two types of SN\,Ia progenitors (short-lived and long-lived), made by 
\citet{Mannuccietal2006}, opens the opportunity to explain not only the overall gradient, but the 
formation of fine structure in radial distribution over the galactic disc for various heavy 
elements by means of a unified mechanism -- the influence of the corotation resonance.

\section*{Acknowledgments}

Authors are grateful to anonymous referee for very important comments.

BMS, AVT and DSW thank Russian Funds for basic research, grants no. 07-02-00454, 08-02-00371 and 
4354.2008.2.


\label{lastpage}

\end{document}